# What Fidelity Metrics Miss: A Structural Check on Synthetic Educational Data

**Hitoshi INOUE[a]* & Koichi YASUTAKE[b]**
[a]Nakamura Gakuen University, Japan
[b]Hiroshima University, Japan
*jinoue@nakamura-u.ac.jp

**Abstract:** Secondary use of educational records is increasingly mediated by platforms that share a differentially private synthetic version of a dataset and validate specific findings against the real data on request. The synthetic version is evaluated by comparing summary statistics of each variable, yet reported confirmation rates suggest that such comparisons do not predict which findings survive. We propose a structural check: the number of connected components of a weekly proximity graph over learners, tracked across a term. Across four annual cohorts of lower-secondary study-habit logs, the synthetic versions reproduced the level of this quantity and the shape of the weekly partition, but its variation across the term was between 2.6 and 4.9 times smaller than in the real data at a common working point, without exception, and those changes fell in different weeks: the synthetic cohorts single out the term's examination weeks and the real cohorts do not. We also show that a routine rule for setting the graph threshold makes naive comparisons between two datasets invalid, and illustrate this with an error of our own. The real curves are also distinguishable from marginal-preserving surrogates of themselves in all four cohorts, where three of the four synthetic ones are not, a comparison that needs no real data; these differences trace to what the generator was given.

## 1. Introduction

Secondary use of the data that learning platforms accumulate is constrained by privacy. One response, now implemented at scale, is to share a differentially private synthetic version of a dataset and to allow researchers to validate specific findings against the real data on request. ReLEAF (Ito, Hsu, & Ogata, 2026c) is such a platform: researchers develop analyses on synthetic cohorts and submit code to be run against the real logs.

The arrangement rests on an assumption about the synthetic data. Exploratory work is done on it, and only conclusions the researcher already believes are put forward for validation. If the synthetic version misrepresents the phenomenon, the researcher does not merely lose time. They may never formulate the question that the real data would have answered. The platform's own pilot found this to be a live concern: across twenty-five validation requests from four researchers, on average 36% of findings obtained on synthetic data were confirmed on the real data, and cohorts with similar statistical fidelity scores produced markedly different confirmation rates (Ito et al., 2026b).

That last observation is the one we take up. Fidelity metrics in common use compare summary statistics of each variable between the real and synthetic versions. Note what that comparison does not include. Whether the weeks of a term differ from one another as the real weeks do is not among the quantities being matched, and a synthetic dataset can match every marginal while having no temporal structure at all. Whether that omission matters is a question about the analyses a dataset will be put to, and the field has argued for some years that learning analytics needs methods that follow learning over time (Knight et al., 2017). We do not settle that question here.

We ask a prior one: whether the omission is consequential in practice. Do real and synthetic cohorts actually differ along a dimension that the metrics do not compare?

We propose checking for exactly that, using a coarse description of how a cohort divides in each week: the number of connected components of a proximity graph over learners, written $\beta_0$, tracked across the weeks of a term. The computation requires only a synthetic dataset and permutations of itself: no case study, no researcher panel, and no access to the real data.

Applying it to four annual cohorts of lower-secondary school study-habit logs, we find that the weekly $\beta_0$ curve of the synthetic version varies between 2.6 and 4.9 times less than that of the real version, with no exception.

This paper makes four contributions. First, we describe a cohort by how it divides in a week, and we follow that description across weeks. Neither the fidelity metrics in use nor the dimensions recently proposed for synthetic sequential data compare data at this level. Second, we show with four annual cohorts that the real and synthetic versions differ at this level, although they agree on the level and shape of the weekly partition. We also show which disclosed statistics constrain the difference and which do not. Third, we run a control experiment on synthetic data whose construction we know. It shows that the check is calibrated, that it can be passed, and that separability from a dataset's own marginals is not a measure of fidelity. Fourth, we explain why a routine parameter-selection rule makes naive comparisons between two datasets invalid, using an error of our own as the example.

## 2. Background

### 2.1 Sharing educational data through synthesis and validation

Differentially private synthetic data generation has been proposed repeatedly as a route to sharing educational records, but the datasets that learning-process research needs are typically small and high-dimensional, a regime in which deep generative models are both costly to build and unreliable (Ito et al., 2026b). The two-stage design that ReLEAF implements avoids the difficulty differently. Synthesis is training-free: differentially private summary statistics are released and used to prompt language models to write a generator. The fidelity that synthesis cannot guarantee is then recovered on demand, through non-private validation of specific analyses against the real data. Requests are reviewed before execution and outputs are restricted to aggregates.

### 2.2 What utility metrics measure

Evaluations of synthetic data in this setting report divergence between the real and synthetic distributions of a fixed set of summary statistics. Ito et al. (2026b) use the average Jensen-Shannon divergence over the median, mean, standard deviation, variance, maximum and root mean square of each time window, and find their training-free method comparable to a deep-learning baseline trained on the real data. Alongside it they introduce epistemic precision, the proportion of a researcher's synthetic findings later confirmed on real data, and report that the two do not track each other.

The gap has a plain reading. A synthetic cohort that matches those twenty-four summary statistics is indistinguishable, on that evidence, from one produced by a generator that never saw how the real data's variables move together. Epistemic precision does reach that question, but only after researchers have worked with the data, formed findings and submitted them, which is late, subjective and expensive.

Two lines of research are relevant here. First, topological summaries have been proposed for evaluating generative models. Charlier et al. (2019) compare the persistence diagrams of a real and a reconstructed distribution using the bottleneck distance. Kim et al. (2023) estimate the support of each distribution and keep only the features that are topologically and statistically significant. Yinzhu Jin et al. (2024) show that generators which map a latent Gaussian into the data space fail on distributions whose topology is not that of a Euclidean space. Each of these studies compares a real dataset with a synthetic one at a single point in time. Second, Kwon et al. (2026) argue that temporal fidelity cannot be inferred from static fidelity, and they define four dimensions for evaluating synthetic sequential data.

The dimension that applies to data of our shape is cross-sectional fidelity, which compares distributions at aligned time points. The authors classify its metrics as working on one feature at a time or on one pair at a time. How a population divides at a given time is not among the quantities that are compared in any of the four dimensions.

### *2.3 Cohort structure as a weekly quantity*

Counting the connected components of a proximity graph over learners gives a coarse summary of how a cohort divides in a given week: high when learners' weekly profiles are mutually dissimilar, low when they concentrate. Tracked across weeks it describes reorganisation rather than level, which is what the fidelity metrics above do not reach.

A closely related construction appears in our earlier work on university virtual-learning-environment logs, where components were counted on the union of two adjacent weeks rather than on a single week (Inoue & Yasutake, 2026). The construction is in turn closely related to established clustering practice: the components of an ε-neighbourhood graph are exactly the clusters of single-linkage clustering cut at distance ε (Gower & Ross, 1969), and clustering learners through a similarity graph built from their trace data is a recognised approach in this field (Peach et al., 2019). The count is also the zeroth Betti number of the associated complex, and topological data analysis has been proposed for learning analytics (Munch, 2017); we use only the count at a single scale and do not compute persistence.

We should be clear about the role this quantity plays here. We do not argue that it is a good measure of learning, nor do we interpret its level educationally. We use it as a structural summary coarse enough to be computed identically on a real cohort and on a synthetic one, and rich enough to change from week to week when the cohort does.

Three properties recommend it over the other descriptors one might track. It is determined by the graph alone, with no optimisation and no random seed, where maximising modularity requires both and would differ between two runs on the same data. It is defined when the graph is disconnected, where mean path length is not, and these graphs are heavily disconnected, with between a seventh and a third of learners alone in a component across our cohorts. And it is a property of the partition rather than a count of edges or triangles, where density and the clustering coefficient are the latter; our question is whether a cohort divides differently from one week to the next, not whether local connectivity changes.

The construction carries a parameter, and the rule normally used to set it has a property that undermines the very comparison we want to make. We return to this in Section 3.2.

## 3. Method

### *3.1 Structural indicator*

For each week t we represent learners as points in a feature space and connect two learners whose Euclidean distance falls below a threshold ε. We write $\beta_0$ for the number of connected components of the resulting graph and read it as the number of distinct behavioural groups the cohort forms that week. We compute it with a connected-components routine. We do not compute persistence and do not use a persistence library.

Because a count alone does not describe how a cohort is divided, we also record the isolated fraction, the proportion of learners whose component has size one, and the component concentration, the sum over components of the squared share of learners each contains. Concentration is the probability that two learners drawn at random from the same week share a component. It is a sum over all components, involves no maximum, and varies continuously as the partition changes. We do not report the size of the largest component. The platform's disclosure rules do not permit component-wise maxima, and the available substitute is the fraction of learners in a majority component, that is, one holding at least half the cohort. That fraction is zero in weeks where no such component exists, so its mean across weeks mixes two regimes, as happened in one of our cohorts.

### 3.2 Choice of working point, and why it must be controlled

The threshold ε must be chosen, and the choice is consequential in a way that is easy to overlook. Both versions of each dataset are analysed under the same primary rule: search a fixed grid from 0.02 to 2.00 in steps of 0.02, and take the ε whose mean weekly $\beta_0$ is closest to a quarter of the cohort size. That target is a convention, not a derived value.

Two consequences follow. First, fixing the mean $\beta_0$ at a quarter of N fixes the number of components but not their sizes: whether a cohort divides into many comparable groups or into one large component with a fringe of singletons is a property of the data, not of the rule. This is why we report partition descriptors alongside $\beta_0$ itself.

Second, and more importantly for what follows, the rule makes the ratio of $\beta_0$ to N approximately one quarter by construction in every dataset to which it is applied. Two datasets analysed each under its own rule-selected ε are therefore not comparable, however natural the comparison looks. We encountered this concretely: an earlier round of our own analysis put each side at its own threshold, 0.68 against 0.70 in one year but 0.36 against 0.68 in another, and the contrast reversed once the quantities were recomputed at a common threshold. We therefore report every cross-dataset quantity at a single common working point, ε = 0.68, fixed in advance from the synthetic cohorts alone as the grid point minimising the summed absolute deviation of $\beta_0/N$ from one quarter; no real data entered its selection. To confirm that our conclusions do not depend on where the threshold sits, we also report all quantities at a pre-specified, data-independent grid of 0.40, 0.60, 0.80 and 1.00.

### 3.3 Data

We use study-habit logs of Japanese lower-secondary school learners obtained through ReLEAF (Ito et al., 2026c), a platform for trustworthy secondary use in which analysis code is executed against the data by the provider and only aggregate statistics are returned. For each annual cohort the platform supplies a synthetic version, generated under differential privacy from summary statistics of the logs, together with on-demand validation against the logs themselves. We analyse both under identical code.

Features are weekly study time in six time-of-day bands: overnight (00:00-04:59), early morning (05:00-08:59), late morning (09:00-11:59), afternoon (12:00-16:59), evening (17:00-20:59) and night (21:00-23:59). Study time is estimated by the provider from the operation logs of a digital textbook system as the elapsed time between the first and last log within each hour, summed into bands and aggregated by week. We analyse four annual cohorts, 2022 to 2025, of 117 to 120 learners. Each cohort covers eighteen ISO weeks, weeks 16 to 28 and 35 to 39, with a six-week gap corresponding to the summer break; the eighteen weeks are therefore not contiguous, and one of the seventeen consecutive-week differences spans seven weeks rather than one. Data are delivered on a complete learner-by-week grid, so a learner with no recorded study in a week is present as a row of zeros. All features are z-standardised globally across the whole period, not within each week, so that weekly graphs share a common scale.

The operational data we received differ in shape from the three cohorts of seventeen weeks and four time-of-day windows described in the platform's own report. We describe the data as delivered and do not speculate about the reason.

For the control experiment in Section 4.5 we also use four presentations of the Open University Learning Analytics Dataset (Kuzilek et al., 2017). We chose them because their cohort sizes differ by a factor of five, from 376 to 2,049 learners over thirty-nine weeks. The features are the weekly click counts on the six most active activity types, and the learners are placed on the same complete learner-by-week grid. This dataset appears here only as a source of real data from which we can build synthetic versions whose construction we know. We do not compare its values with the values of the study-habit cohorts.

### 3.4 What we compare

Our primary comparison is the coefficient of variation of the weekly $\beta_0$ curve, computed at the common working point. This asks whether a synthetic cohort reproduces not the level of $\beta_0$, which the selection rule largely fixes, but its movement across weeks. We use the coefficient of variation rather than the variance because mean $\beta_0$ differs across cohorts by more than a factor of two and a scale-free measure is needed to compare them, and rather than the range because the platform's disclosure rules do not permit returning extreme values. A robust scale-free alternative, such as the median absolute deviation over the median, would serve as well; we use the coefficient of variation for continuity with Section 3.5.

We also ask whether each version can be separated from surrogates that preserve its own within-week marginals. We construct the surrogates by permuting each feature column independently within each week, which preserves the marginal distribution of every week exactly. We generate 200 surrogates, which resolves the reported proportion to 0.005. We compare $\beta_0$ curves by the mean absolute difference, and we treat the observed curve and the surrogate curves in the same way.

Let $d_{obs}$ be the mean distance from the observed curve to the surrogates, and let $d_i$ be the mean distance from surrogate i to the other surrogates. We report the proportion of surrogates for which $d_i$ is at least $d_{obs}$, together with the standardised position of $d_{obs}$ in the distribution of $d_i$. Under the hypothesis that $\beta_0$ depends on the data only through the weekly marginals, the observed curve is exchangeable with the surrogates, so this is a valid permutation test. A large proportion means that the observed curve lies inside the surrogate swarm.

Note what the surrogate destroys. Permuting columns within a week removes association between learners, but it also removes association among the six bands of a single learner, in particular the co-occurrence of zeros when a learner does not study at all that week. A small proportion therefore establishes structure beyond independent within-week marginals. It does not by itself identify that structure as holding between learners.

Finally we compare the isolated fraction and the component concentration at the common working point, and the per-learner Spearman correlation between the proportion of weeks a learner spends in the majority component and that learner's total study time. This last analysis uses no outcome labels.

### *3.5 Calibrating the interpretation*

Because the permutation test does not identify the structure it detects, we calibrate against two mechanisms that could produce a small proportion, injecting each into a synthetic cohort and re-running the identical procedure. Making an additional k learners per week all-zero raises each band's marginal zero rate, which the surrogate also reproduces, so only the co-occurrence is detected; this moves the proportion from 0.800 with no injection, at 8.8 all-zero learners per week, through 0.015 at k = 20 and 28.8 learners, to 0.000 at k = 30 and 38.8 learners. Scaling all values in a week by a common factor moves the coefficient of variation of the weekly mean study time and that of $\beta_0$ in near lock-step, from 0.181 and 0.149 through 0.448 and 0.466 to 0.744 and 0.761. Both calibrations let us ask, of a real cohort, whether its measured all-zero rate and its measured activity variation are large enough to account for what the test reports.

## 4. Results

### *4.1 The level and shape of the weekly partition are reproduced*

At the common working point, the synthetic and real versions of each cohort divide their learners in broadly comparable ways. Table 1 reports the mean weekly $\beta_0$ relative to cohort size, the proportion of learners alone in a component, and the probability that two learners drawn from the same week share a component.

Two of the four cohorts agree to within a few percent on every column. The other two differ more: in 2023 the synthetic version is the less fragmented of the pair by 55%, and in 2024 by 24%. On these quantities alone, a reader would have no reason to doubt the synthetic cohorts in at least half the years.

*Table 1. Partition descriptors at the common working point.*

| Cohort | $\beta_0$/N (synthetic) | $\beta_0$/N (real) | isolated (synthetic) | isolated (real) | concentration (synthetic) | concentration (real) |
|---|---|---|---|---|---|---|
| 2022 | 0.245 | 0.258 | 0.189 | 0.222 | 0.448 | 0.529 |
| 2023 | 0.163 | 0.252 | 0.138 | 0.216 | 0.685 | 0.494 |
| 2024 | 0.235 | 0.291 | 0.210 | 0.254 | 0.569 | 0.461 |
| 2025 | 0.377 | 0.399 | 0.327 | 0.355 | 0.324 | 0.352 |

The association between component membership and study volume is likewise reproduced. Correlating, across learners, the proportion of weeks a learner spends in the majority component against that learner's total study time gives a negative coefficient in every cohort and every version, from -0.97 to -0.81 in the synthetic data and from -0.90 to -0.63 in the real data. The synthetic values are the stronger of each pair in three cohorts of four, but the sign and the order of magnitude carry over.

### *4.2 The temporal variation is not*

The picture changes when we look at how much these quantities move across the term. Table 2 reports the coefficient of variation of the weekly $\beta_0$ curve, at the same working point.

*Table 2. Variation of the weekly $\beta_0$ curve at the common working point.*

| Cohort | CV (synthetic) | CV (real) | ratio |
|---|---|---|---|
| 2022 | 0.149 | 0.539 | 3.6 |
| 2023 | 0.152 | 0.399 | 2.6 |
| 2024 | 0.086 | 0.423 | 4.9 |
| 2025 | 0.126 | 0.431 | 3.4 |

The two ranges do not overlap. In every cohort the real curve moves between 2.6 and 4.9 times as much across the term as the synthetic curve. In addition, the cohort with the closest agreement in Table 1 is not the cohort with the closest agreement here. The gap is not a difference of degree between similar objects. A curve whose coefficient of variation is 0.15 or below is close to flat, and the partition that it describes is close to unchanging from week to week.

This result does not depend on the position of the threshold. At the pre-specified reference grid the same separation holds at all four values. The synthetic coefficients are between 0.05 and 0.17 and the real ones are between 0.34 and 0.54. Across the four cohorts and the four thresholds we have sixteen synthetic values against sixteen real values, and no synthetic value reaches any real value.

### *4.3 What a marginal-preserving surrogate recovers*

We next ask whether each version can be told apart from surrogates preserving its own within-week marginals. At the common working point every real cohort separates: no surrogate lies further from the swarm than the observed curve, in all four years, with standardised positions of 3.77, 3.87, 6.36 and 7.81. Three of the four synthetic cohorts do not, returning 0.800, 0.05 and 0.09 at positions of -0.92, 1.78 and 1.47; the 2025 synthetic cohort separates at 3.70.

Section 3.5 asked what a measured all-zero rate could account for: 8.8 all-zero learners per week left the proportion at 0.800, and 28.8 were needed to reach 0.015. The real cohorts carry 18.2, 7.4, 2.1 and 4.4. The 2024 cohort thus separates at 6.36 with fewer all-zero learners than the uninjected calibration, which co-occurring zeros cannot explain; the 2022 cohort, at 18.2, sits where they could account for part of it. Removing all-zero rows leaves the real $\beta_0$ curves unchanged to three significant figures and moves the positions to 4.08, 4.21, 6.44 and 7.73, so the separation is not an artefact of those rows; the same operation weakens the synthetic 2024 result from 0.09 to 0.225.

One caution governs how these may be compared. Each real cohort's own rule-selected threshold, reported in Section 3.5 for reference only, sits 0.02, 0.00, 0.10 and 0.32 above the common one, and the two furthest from it are the two with the largest standardised positions. With four cohorts this is an observation rather than a demonstrated relationship, but it runs in the direction Section 3.2 anticipates. We therefore read the test cohort by cohort and do not compare its magnitude across cohorts. The coefficient of variation shows no such ordering, which is one reason we treat it as primary.

### *4.4 Where the difference comes from*

The second calibration of Section 3.5 can now be read against the real cohorts. The measured variation of weekly mean study time is 0.33 to 0.61 in the real cohorts and 0.12 to 0.18 in the synthetic ones, and across all eight datasets the variation of $\beta_0$ lies between 0.73 and 1.27 times it. The calibration therefore shows that week-to-week movement in the level of activity is sufficient to produce a gap of the size Table 2 reports, and the real cohorts carry movement of that size. We read the gap as consistent with that mechanism rather than as demonstrating that nothing else contributes.

Why that movement is absent follows from what is disclosed. For the study-habit records the differentially private summary comprises the proportion of zero values across the 108 band-by-week combinations, an entropy over bands computed from each band's eighteen-week total, and an entropy over weeks computed from each week's total across bands. The second marginalises over weeks and the third over bands. No released statistic constrains which band and which week are zero together. Both consequences are visible. The marginal zero rate, which is released, is reproduced to within 3% to 13%. The excess of all-six-band zeros over what independent bands would give is not: it exceeds one in every real cohort, from 1.40 to 2.57, while the synthetic values run from 0.41 to 2.15 and fall below one in two cohorts of four.

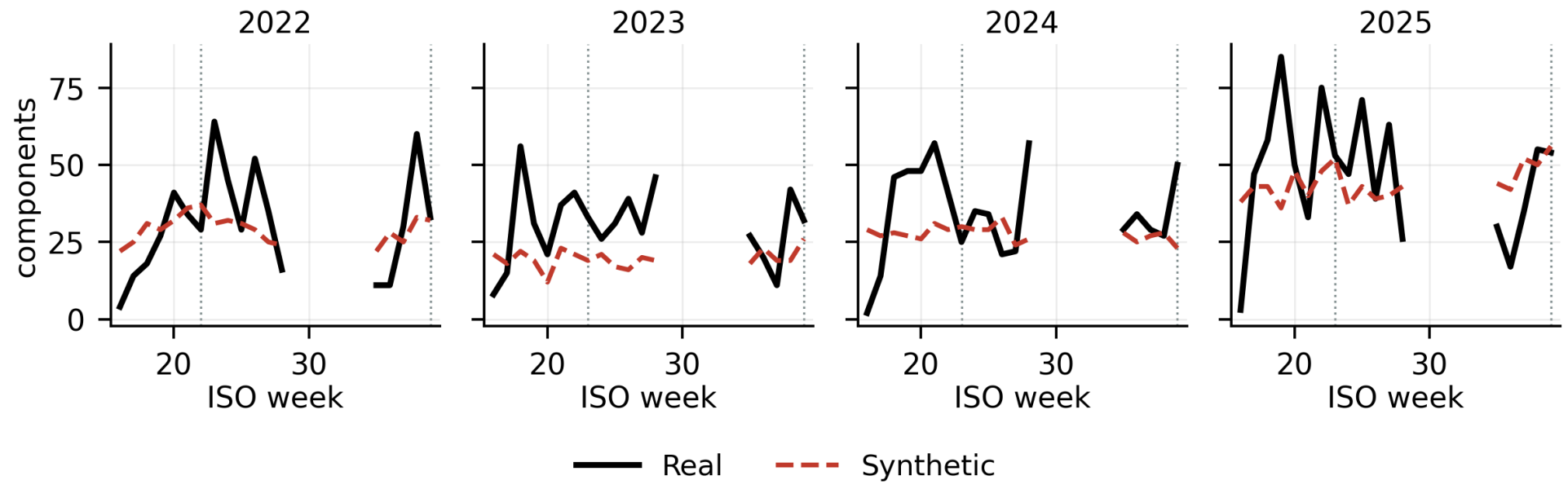

*Figure 1. Weekly component counts at the common working point. Dotted lines mark the examination weeks; the gap is the summer break.*

An entropy over weeks is invariant to permuting them, so the summary carries no information about which week is which. The generator has it from elsewhere: the published procedure supplies a description of the data alongside the private summary, and the dataset's documentation records the term's examination schedule. We do not see the prompt, but the synthetic cohorts track a year-specific detail of it. The mid-term examination falls in week 22 in 2022 and in week 23 in the other three years, and in every cohort the synthetic mean study time is higher in that year's own examination week, and among the two highest of the eighteen.

The real cohorts are not: there the mid-term week ranks ninth to fifteenth, and in two cohorts falls below the mean of the rest. The synthetic cohorts therefore miss not only the size of the week-to-week movement but also its placement, and the Spearman rank correlation between the real and synthetic weekly $\beta_0$ curves runs from -0.22 to 0.57. Figure 1 shows the eight curves. Why the real cohorts study less in examination weeks is a question our data cannot answer, and we do not speculate.

*4.5 Can the check be passed?*

The results so far show that one generator's output fails a check. They do not show that the check can be passed. If every synthetic dataset separated from its own surrogates and understated the variation of $\beta_0$, the check would be of no use in telling generators apart. We therefore ran it on synthetic data of known construction.

From the four OULAD presentations of Section 3.3 we produced three synthetic versions of each. The first permutes each column of the learner-by-week matrix independently, preserving every marginal and retaining no dependence; because this is the operation the surrogate itself performs, its output is a draw from the null and its test statistic should be uniformly distributed. The second samples from a multivariate normal matched to the mean and covariance. The third is a Gaussian copula: empirical marginals with the rank correlation of the full learner trajectory preserved. All are compared with the real data at the threshold selected on the real data, under a single standardisation shared across conditions.

Three things follow. First, the check is calibrated rather than merely permissive. Over twenty-four replications the column-permuted version produced test statistics spread across the interval, from 0.010 to 0.860 with a mean of 0.537, and a test of uniformity does not reject. One of the twenty-four fell below 0.05, against an expectation of 1.2 under uniformity.

Second, the check can be passed. Only the copula reproduces the weekly values themselves: its coefficient of variation lands within 3.5% of the real one in all four presentations, with mean $\beta_0$ within 6%. Dividing each synthetic curve by the real one, the copula gives a nearly constant ratio, with residuals of 3% to 4% of the mean level after a single scale factor, where the column-permuted version leaves 22% to 26% and the multivariate normal 18% to 36%. Preserving the weekly marginals recovers the descent the four series share. Preserving the rank correlation of the learner trajectory recovers the rest.

Third, and least expected, separability is not fidelity. The multivariate normal version separated from its own surrogates more sharply than the real data did in every presentation, by as much as a factor of two, while misstating the level of $\beta_0$ by between 28% and 143% and its variation by between 14% and 38%. A synthetic dataset can be highly distinguishable from its own marginals and a poor description of the cohort.

## 5. Discussion

Because these platforms ask researchers to explore on synthetic data and to bring forward only formed conclusions, what is not noticed on the synthetic version is not merely unconfirmed but unasked. We can say what a researcher working on these synthetic cohorts would see: a cohort whose division into groups is close to unchanging across the term.
What follows from that depends on the analysis. For questions about who differs from whom in a given week, the synthetic data reproduces what is needed. For questions about how much a cohort reorganises, it does not, and a researcher would have little in the data to prompt such a question in the first place. A concrete case: an analysis asking whether a cohort was structurally different in the weeks around an examination, or before and after some change in a course, would find almost no week-to-week difference to report on the synthetic data, and would find a large one on the real data. The finding would not fail validation; it would never be submitted. Omissions of this kind are one possible reason why the reported rate of confirmed findings is low and does not track statistical fidelity. We have not tested that link, and nothing below rests on it.

Section 4.4 traces the gap to what is disclosed: two of the three released study-habit statistics are marginalisations of the same band-by-week matrix, one over weeks and one over bands, and neither constrains how one week differs from another or how the weeks of different learners align. Whether a summary that does constrain those things can be released under differential privacy, at a cost in utility that leaves the rest of the analysis intact, is a separate question we do not address.

As a practical matter the check is cheap and early. It needs a synthetic dataset and permutations of itself, so a custodian can run it before release and a researcher before committing to a line of analysis. Without real data, that run yields the surrogate comparison of Section 4.3, which flagged three of these four synthetic cohorts; reading the size of the gap, as in Table 2, needs a real reference. Where epistemic precision reaches the same question only after a study has been conducted and its findings submitted, this reaches a part of it beforehand, at the cost of examining one property rather than the researcher's actual conclusions.

A word on what we have and have not claimed for the quantity itself. Our earlier work was motivated by two questions: the non-stationarity of learning processes, and the description of a cohort at the group rather than the individual level. These are the questions for which a weekly structural summary would be wanted. We have not argued here that $\beta_0$ answers them well, and nothing in this paper depends on its doing so. Which structural descriptor serves this purpose best is a separate question we leave open. The properties set out in Section 2.3 are reasons to prefer $\beta_0$ here, not evidence that alternatives would fail. What the paper shows is narrower and, we think, more robust for being narrow: along a dimension that current fidelity metrics do not compare, real and synthetic cohorts differ systematically, and part of the difference can be detected before any real data is touched.

## 6. Limitations

Seven limitations bound what the above can be taken to show.

First, as noted in Section 3.4, the surrogate removes association between learners and association among a learner's own bands at the same time, so a separation from the surrogate swarm does not identify which of the two is responsible. Our calibration in Section 3.5 bounds how much of it the co-occurrence of zeros could account for, but does not settle the question.

Second, features are standardised using each dataset's own dispersion. A generator that misstates the spread of a band is not penalised by any quantity we report. What we compare is shape after standardisation. The dispersions we observed differ substantially between cohorts, so this is not a hypothetical exclusion.

Third, we cannot separate differences between years from differences between generation conditions. The platform's multi-shot design (Ito et al., 2026a) feeds researchers' feedback from one cycle into the generation of the next, and which feedback is reflected in each of the four synthetic cohorts we analysed is not determinable from public information.

Fourth, the removal of all-zero rows was carried out only within the scope of an approved validation request, and the corresponding re-selection of the threshold on the filtered data is reported for reference rather than used as a second working point.

Fifth, a zero in these features means that no interval of study time could be measured in that band and week, not that the learner did not study: the provider's estimator takes the elapsed time between the first and last operation log within each hour, so a single log leaves a zero. Our co-occurrence result concerns the pattern of these zeros, but the level of the zero rate should not be read as a rate of non-study.

Sixth, the platform does not warrant the empirical validity of its synthetic cohorts, and none of the values we report for them should be read as describing the learners they stand for.

Seventh, the evidence comes from one school subject at one institution, four cohorts of 117 to 120 learners over eighteen weeks each, delivered through one platform. We do not claim that the size of the gap generalises. The check itself transfers to any dataset of this shape, but establishing that it is informative elsewhere would require further datasets and further synthesis mechanisms.

## 7. Conclusion

Synthetic educational data are evaluated, in practice, by comparing summary statistics of each variable between the real and synthetic versions. We asked what such a comparison does not see, using the number of connected components of a weekly proximity graph over learners as a description of how a cohort divides.

Across four annual cohorts of lower-secondary study-habit logs shared through a two-stage platform, the synthetic versions reproduced the level of this quantity and the shape of the weekly partition, but not its movement: at a common working point the real curves varied between 2.6 and 4.9 times as much across the term, without exception. A researcher working on the synthetic data would see a cohort whose structure barely changes from week to week, and would see what change there is placed in weeks where the real cohort does not place it.

Computing the quantity costs one synthetic dataset and permutations of itself. That much can be done before any real-data validation is requested, though measuring the size of the gap still needs the real data. What that comparison reaches is a property that the summary-statistic comparisons do not.

## Acknowledgements

This work was supported by JSPS KAKENHI Grant Numbers JP25K00845, JP25K21951, JP26K00528, and JP26K22187. We thank the ReLEAF team at Kyoto University for access to the platform and for executing our validation requests against the real data.